\documentclass[a4paper,11pt]{article} 
  
\usepackage[english]{babel}
\usepackage[a4paper]{geometry}
\usepackage{amsmath}
\usepackage{amsthm}
\usepackage{amssymb}
\usepackage{color}

\usepackage{hyperref}         

\def\bR{\mathbb{R}}

\def\bN{\mathbb{N}}

\def\cC{\mathcal{C}}

\def\cQ{\mathcal{Q}}

\def\cM{\mathcal{M}}

\def\cV{\mathcal{V}}

\def\cF{\mathcal{F}}
\def\cG{\mathcal{G}}
\def\cL{\mathcal{L}}
\def\cJ{\mathcal{J}}
\def\cN{\mathcal{N}}
\def\cE{\mathcal{E}}

\def\cH{\mathcal{H}}
\def\cR{\mathcal{R}}

\def\ph{\varphi}

\def\ZZZ{\mathbb{Z}}

\def\indic{\hbox{\raise-2pt \hbox{\indbf 1}}}

\let\==\equiv

\let\0=\noindent

\def\*{{\hfill\break\null\hfill\break}}

\def\tende#1{\,\vtop{\ialign{##\crcr\rightarrowfill\crcr
             \noalign{\kern-1pt\nointerlineskip}
             \hskip3.pt${\scriptstyle #1}$\hskip3.pt\crcr}}\,}
\def\otto{\,{\kern-1.truept\leftarrow\kern-5.truept\to\kern-1.truept}\,}

\def\tr{{\rm tr}}

\newtheorem{theorem}{Theorem}[section]  

\numberwithin{equation}{section}

\def\be{\begin{equation}}
\def\ee{\end{equation}}

\let\a=\alpha \let\b=\beta

          \let\ph=\varphi   
        
        \let\L=\Lambda

\begin{document}

\title{The Excitation Spectrum of the Bose Gas in the Gross--Pitaevskii Regime} 

\author{Chiara Boccato\\
\\
IST Austria, Am Campus 1\\
3400 Klosterneuburg,
Austria\\
}

\date{}

\maketitle

\begin{abstract}	
We consider a gas of interacting bosons trapped in a box of side length one in the Gross--Pitaevskii limit. We review the proof of the validity of Bogoliubov's prediction for the ground state energy and the low--energy excitation spectrum. This note is based on joint work with C. Brennecke, S. Cenatiempo and B. Schlein.
\end{abstract}

\section{Introduction}

The first theoretical investigation of the Bose gas dates back to 1924: in \cite{B24,E24} Bose-Einstein condensation was predicted for the non--interacting Bose gas.  The interacting problem is considerably more difficult. An important progress on the study of the interacting Bose gas has been made by Bogoliubov \cite{B} in 1947.
Bogoliubov's heuristic approach leads to  expressions for the ground state energy and the excitation spectrum, the latter explaining the superfluid behavior of substances such as liquid helium.
Bogoliubov's method sets the track for approaching the study of the interacting Bose gas; 
however, the physical intuitions behind it require a
precise mathematical formulation.
 In this note we present a rigorous implementation of Bogoliubov theory for the interacting Bose gas in the Gross--Pitaevskii regime.

We consider a gas of $N$  bosons trapped in a box $\Lambda=[-1/2,1/2]^3$ with periodic boundary conditions. The Hamiltonian $H_N$, given by
\begin{equation}\label{Hxspace}
 H_N=-\sum_{i=1}^{N}\Delta_{x_i}+N^2\sum_{i< j}^NV(N(x_i-x_j)),
\end{equation}
acts on the space of symmetric (bosonic) wave functions
\[
 L^2_s(\L^{N})=\{\psi\in L^2(\L^{N}): \psi(x_{\sigma(1)},\dots,  x_{\sigma(N)})=\psi(x_1,\dots,x_N), \text{ for every }\sigma\in S_N \},
\]
where $S_N$ is the set of all permutations of $N$ objects.  We take the interaction potential $V$ to be non--negative, spherically symmetric and of finite range.
For large $N$, the scaling of the interaction potential models a very dilute system: the range of interactions is of order $N^{-1}$, while the mean interparticle distance is much larger, namely of order $N^{-1/3}$; this is called  the Gross--Pitaevskii regime. Equivalently we can describe the system as trapped in a box of side length $N$ with non--rescaled interactions. In the latter picture the density is $N^{-2}$ and it is clear that the limit of large $N$ is a simultaneous large volume and low density limit.

The ground state energy of this system is known \cite{LY,LSY}, to leading order in $N$, to be
\begin{equation}\label{eq:LSY} 
E_N = 4 \pi \frak{a}  N + o (N) \, .
\end{equation}
In the expression above the \emph{scattering length} $\frak{a}$ appears; it is defined by 
\begin{equation}\label{eq:sl}
  8 \pi \frak{a}  =  \int_{\mathbb{R}^3} V(x) f(x) dx,
\end{equation}
where $f$ is the solution of the zero--energy scattering equation
\begin{equation}\label{zeroEnergySE} \left[ - \Delta + \frac{1}{2} V (x)  \right] f (x) = 0 \end{equation}
with the boundary condition $f (x) \to 1$, as $|x| \to \infty$.

The ground state vector \cite{LS}, and any sequence of approximate ground state vectors  \cite{LS2,NRS}, i.e., any sequence $\psi_N \in L^2_s (\L^N)$ with $\| \psi_N \| = 1$ and  
\begin{equation}\label{eq:appro-gs} 
\lim_{N \to \infty} \frac{1}{N} \langle \psi_N , H_N \psi_N \rangle = 4 \pi \frak{a} \, , 
\end{equation}
exhibit \emph{Bose--Einstein condensation}. This means that the reduced density matrices $\gamma_N = \tr_{2, \dots , N} |\psi_N \rangle \langle \psi_N |$ satisfy, as $N\to\infty$,
\begin{equation}\label{eq:BEC0} 
 \langle \varphi_0 , \gamma_N \varphi_0 \rangle \to 1
\end{equation}
where $\ph_0 \in L^2 (\Lambda)$ is the one--particle zero momentum mode $\ph_0 (x) = 1$, called the condensate wave function. The expectation on the left--hand side of equation \eqref{eq:BEC0} is the fraction of particles in the zero momentum mode; equation \eqref{eq:BEC0} establishes that all particles, up to a fraction vanishing in the limit $N\to\infty$ are in the condensate state.

We present now our results obtained in \cite{BBCS3,BBCS4}.
In Theorem \ref{thm:mainCond} below we determine the convergence rate of condensation, improving \eqref{eq:BEC0}.  Theorem \ref{thm:main}   exhibits the next order to \eqref{eq:LSY} for the ground state energy and determines the excitation spectrum of \eqref{Hxspace}.


\begin{theorem}[Optimal rate for Bose--Einstein condensation]\label{thm:mainCond}
Let $V \in L^3 (\mathbb{R}^3)$ be non--negative, spherically symmetric and compactly supported.
Let $\psi_N \in L^2_s (\Lambda^N)$ be a sequence with $\| \psi_N \| = 1$ such that   
\begin{equation}\label{eq:appGS}
  \langle \psi_N , H_N \psi_N \rangle  \leq 4 \pi \frak{a} N + \zeta
\end{equation}
for a $\zeta > 0$. Then the reduced density matrix associated with $\psi_N$ satisfies
\begin{equation}\label{eq:BEC} 1 - \langle \varphi_0 , \gamma_N \varphi_0 \rangle \leq  \frac{C(\zeta+1)}{N} \end{equation}
for all $N \in \mathbb{N}$ large enough.
\end{theorem}

Equation \eqref{eq:BEC} establishes a bound, uniform in $N$, for the number of excited particles over the condensate. This holds for the ground state vector and for approximate ground states. The proof of  Theorem \ref{thm:mainCond} also provides the estimate
 \begin{equation}\label{eq:wrongConst}
    |E_N-4\pi \frak{a}N|\leq D
 \end{equation}
for a $D>0$, improving \eqref{eq:LSY}. Results \eqref{eq:BEC} and \eqref{eq:wrongConst} have been obtained in \cite{BBCS4} (and before in \cite{BBCS1} for small interaction potentials). Going beyond \eqref{eq:BEC} and \eqref{eq:wrongConst} requires additional techniques, which we developed in \cite{BBCS3}. In the theorem below we state our result for the second order of the ground state energy and for the energy of excitations.

\begin{theorem}[Ground state energy and excitation spectrum]\label{thm:main}
Let $V$ be as in Theorem \ref{thm:mainCond}. Then, for $N \to \infty$, the ground state energy is given by 
\begin{equation}
\begin{split}\label{1.groundstate}
E_{N} = \; &4\pi (N-1) \frak{a}+ e_\Lambda \frak{a}^2 \\ & - \frac{1}{2}\sum_{p\in\Lambda^*_+} \left[ p^2+8\pi \frak{a}  - \sqrt{|p|^4 + 16 \pi \frak{a}  p^2} - \frac{(8\pi \frak{a} )^2}{2p^2}\right] + \mathcal{O} (N^{-1/4}) \, . 
    \end{split}
    \end{equation}      
 Here
 we introduced $\Lambda_+^* = 2\pi \mathbb{Z}^3 \backslash \{0 \}$ and 
 we defined    
 \begin{equation}\nonumber
   e_\Lambda = 2 - \lim_{M \to \infty} \sum_{\substack{p \in \mathbb{Z}^3 \backslash \{ 0 \} : \\ |p_1|, |p_2|, |p_3| \leq M}} \frac{\cos (|p|)}{p^2}. \end{equation}
 Moreover, the spectrum of $H_N-E_{N}$ below a threshold $\zeta$ consists of eigenvalues given, in the limit $N \to \infty$, by 
    \begin{equation}
    \begin{split}\label{1.excitationSpectrum}
    \sum_{p\in\Lambda^*_+} n_p \sqrt{|p|^4+ 16 \pi \frak{a} p^2}+ \mathcal{O} (N^{-1/4} (1+ \zeta^3)) \, . 
    \end{split}
    \end{equation}
Here $n_p \in \mathbb{N}$ for all $p\in\Lambda^*_+$ and $n_p \not = 0$ for finitely many $p\in \Lambda^*_+$ only.  
\end{theorem}

Theorem \ref{thm:main} confirms Bogoliubov's predictions  for the equilibrium properties of the interacting Bose gas. Equation \eqref{1.groundstate} is the finite volume analogue of the well--known Lee--Huang--Yang formula. The sum in the second line of \eqref{1.groundstate} gives a contribution of order one, since the summand behaves as $p^{-4}$ for large $p$. We find also a boundary contribution $e_\Lambda \frak{a}^2$, of order one too, due to the fact that we work in a finite box. The excitation spectrum \eqref{1.excitationSpectrum} is given by a sum of approximately non--interacting harmonic oscillators.
We read there the dispersion relation of excitations $\mathcal{E}(p)=\sqrt{|p|^4+ 16 \pi \frak{a}  p^2}$. For small momenta, $\mathcal{E}(p)=\sqrt{16\pi \frak{a}}|p|\big(1+\mathcal{O}(p^2)\big)$ is linear, in contrast with the quadratic behavior of the dispersion relation of non--interacting particles. In \cite{B} Bogoliubov associated such linear behavior with  the phenomenon of superfluidity, connecting it to Landau's arguments \cite{Lan}.

\section{Bogoliubov theory}

We present here Bogoliubov's approximation procedure \cite{B} for deriving the energy spectrum of a bosonic gas.
Bogoliubov writes the Hamiltonian \eqref{Hxspace} in second quantized form, i.e.,
\begin{equation}\label{eq:Hmom} H_N = \sum_{p \in \Lambda^*} p^2 a_p^* a_p + \frac{1}{2N} \sum_{p,q,r \in \Lambda^*} \widehat{V} (r/N) a_{p+r}^* a_q^* a_{p} a_{q+r}, 
\end{equation}
where $a^*_p$ and $a_p$ are creation and annihilation operators associated with momentum $p\in\L^*=2\pi\ZZZ^3$. We denote with $\widehat V$ the Fourier transform of the interaction potential  $V$. (To be precise, Bogoliubov works in a  thermodynamic limit setting, with a non--rescaled interaction; however, we discuss in this note only the Gross-Pitaevskii regime.) The Bogoliubov approximation consists of three steps.
\begin{itemize}
 \item[a.] Replacing creation and annihilation operators corresponding to the zero-momentum mode $a_0^*,\,a_0$ by the  number $N^{1/2}$ (i.e., imposing condensation by assumption).
 This procedure extracts the condensate contributions and decomposes $H_N$ in a sum of constant (i.e., not containing operators) contributions, quadratic, cubic and quartic contributions in creation and annihilation operators of non--zero modes. (There are no linear terms due to translation invariance.)
\item[b.] Dropping all terms in the  Hamiltonian that are higher than quadratic in $a^*_p$ and $a_p$, for $p\neq 0$. The resulting Hamiltonian is quadratic
and can be explicitly diagonalized through a Bogoliubov transformation. Diagonalization yields the ground state energy and the excitation spectrum. However, the result differs already at leading order \eqref{eq:LSY}. There only the first two summands of the Born series for the scattering length, $\frak{a}^{(0)}=(8\pi)^{-1}\widehat{V}(0)$ and $\frak{a}^{(1)}=-\frac{1}{16\pi} \sum_{p\in\Lambda^*_+ } \frac{\widehat{V} (p)^2}{p^2} $, appear.
In the second order and in the excitation spectrum only the Fourier transform of the interaction $\widehat{V}(p)$ appears.
\item[c.] Substituting in the result the  Born approximations and $\widehat{V}(p)$ with the scattering length $\frak{a}$ leads to  \eqref{1.groundstate} and \eqref{1.excitationSpectrum}.
\end{itemize}

With a rigorous analysis, the inclusion of cubic and quartic terms (neglected in Step b.) should take care of the appearance of the full scattering length in the result (which Bogoliubov introduces by hand in Step c.).
In the next sections we will see how to extract from cubic and quartic terms these order one contributions to the ground state energy and the excitation spectrum.

 Bogoliubov's ideas have been implemented for the derivation of the ground state energy of the Bose gas in the thermodynamic limit (the Lee--Huang--Yang formula) \cite{GiuS,BriS1,BriFS,FS}, and before for the computation of the ground state energy of the bosonic jellium \cite{LSo}.
 For the Bose gas in the mean field regime more information is available, and Bogoliubov's method has been implemented to give the excitation spectrum (see \cite{Sei,GS,LNSS,DN,P3}). In this regime the gas is confined in a fixed volume and the interaction scales as $V_N(\cdot)=N^{-1}V(\cdot)$, describing a high density system. Here collisions between particles are sensitive to the shape of the interaction potential, and the excitation spectrum has the form
     \begin{equation}
    \begin{split}\label{1.excitationSpectrumMF}
    \sum_{p\in\Lambda^*_+} n_p \sqrt{|p|^4+ 2 \widehat{V}(p)  p^2} \, ,
    \end{split}
    \end{equation}
 with $n_p \not = 0$ only for finitely many $p\in \Lambda^*_+$.
 When, instead, we consider dilute, strongly interacting  regimes, correlation effects renormalize interactions and lead to the emergence of the scattering length instead of $\widehat{V}(p)$.
 This is what makes the implementation of Bogoliubov theory in the Gross--Pitaevskii regime more difficult than in the mean field case.

 From a slightly different point of view, one could minimize the Hamiltonian after Step a. over all quasi--free states. This procedure can be viewed as a variational formulation of Bogoliubov theory and gives the leading order of the ground state energy correctly, but the second order is off by a constant (see \cite{ESY,NRS1,NRS2}).
In \cite{YY} the second order is correctly resolved using a trial state which additionally includes correlations that can be thought as arising from cubic combinations of creation and annihilation operators.
 
\section{Bose-Einstein condensation}

We present now the ideas developed in \cite{BBCS4} leading to the proof of Theorem \ref{thm:mainCond}.
Our goal is to obtain a bound, uniform in $N$, for the number of excited particles over the condensate. We denote this quantity by  $\cN_+ = \sum_{p \in \Lambda^*_+} a_p^* a_p$.

\emph{Step 1: Fock space of excitations.} We start with a technique developed in \cite{LNSS}, where it was observed that every wave function $\psi_N\in L^2_s(\L^N)$ can be uniquely decomposed as $\psi_N=\sum_{j=0}^N\psi^{(j)}_N\otimes_s\ph_0^{\otimes(N-j)}$ for a sequence $\psi_N^{(j)} \in L^2_{\perp} (\Lambda)^{\otimes_s j}$. With $L^2_\perp (\Lambda)^{\otimes_s j}$ we indicate the symmetric tensor product of $j$ copies of the orthogonal complement $L^2_{\perp} (\Lambda)$ of $\ph_0$ in $L^2 (\Lambda)$. We take $\ph_0$ to be the condensate wave function $\ph_0(x)=1$. We organize the coefficients $\psi_N^{(j)}$ as a vector in a bosonic Fock space $ \cF^{\leq N}_+ = \bigoplus_{j = 0}^N L^2_\perp (\Lambda)^{\otimes_s j}$ constructed over $L^2_\perp (\Lambda)$ and truncated to sectors with at most $N$ particles. This suggests to define a unitary map $U_N: L^2_s (\Lambda^{N}) \to \cF_{+}^{\leq N} $ through $ U_N \psi_N = \{ \psi^{(0)}_N, \psi^{(1)}_N, \dots , \psi^{(N)}_N  \}$.
The map $U_N$ factors out the Bose--Einstein condensate contribution in $\psi_N$ and returns the excitations.
Using $U_N$ we define a new Hamiltonian
\begin{equation}\label{eq:cL}
 \cL_N = U_N H_N U_N^* : \cF^{\leq N}_+ \to \cF_+^{\leq N}
 \end{equation} 
describing excitations over the condensate. 
Conjugation of $H_N$ with $U_N$ acts as a replacement of  creation and annihilation operators $a_0^*, a_0$ of the zero-momentum mode by $(N-\cN_+)^{1/2}$, while, for $p\neq0$, $a_p$ and $a^*_p$ remain untouched:
\begin{equation}\nonumber
\begin{split} 
U_N a^*_0 a_0 U_N^* &= N- \cN_+  \\
U_N a^*_p a_0 U_N^* &= a^*_p \sqrt{N-\cN_+ } \\
U_N a^*_0 a_p U_N^* &= \sqrt{N-\cN_+ } a_p \\
U_N a^*_p a_q U_N^* &= a^*_p a_q .
\end{split} \end{equation}
This procedure rigorously implements  Bogoliubov's first step. Notice that, in addition, the resulting excitation Hamiltonian, unitarily related to $H_N$, preserves the truncated Fock space  $ \cF^{\leq N}_+$.  The excitation Hamiltonian $ \cL_N$ decomposes into a sum of constant terms (i.e., not containing creation and annihilation operators) and quadratic, cubic and quartic contributions in creation and annihilation operators (up to factors $(N-\cN_+)^{1/2}$). Now the central task is to use the energy bound \eqref{eq:appGS} in the assumptions in order to control
\begin{equation}\nonumber
1 - \langle \varphi_0 , \gamma_N \varphi_0 \rangle =1 -N^{-1} \langle \psi_N, a_0^* a_0 \psi_N \rangle=N^{-1} \langle \psi_N, U_N^* \cN_+ U_N \psi_N \rangle.
\end{equation}
However, there is a difficulty: in $\cL_N$ the constant contribution of order $N$ is $N \widehat{V} (0)/2$, which does not agree with the correct leading order \eqref{eq:LSY} of the ground state energy. Some other important contributions are therefore hidden in the remaining part of $\cL_N$. Those are in fact the contributions of correlations, which are not included in the action of $U_N$.

\emph{Step 2: Generalized Bogoliubov transformation.} To extract the energy of correlations, we conjugate $\cL_N $ further with a generalized Bogoliubov transformation $e^{B(\eta)}$. This transformation is different from the transformation that Bogoliubov used in Step b. to diagonalize his quadratic effective Hamiltonian. Our approach instead is inspired by the treatment of the dynamics in \cite{BDS,BS}; a similar transformation was also used before in \cite{ESY} for the computation of the ground state energy in the thermodynamic limit. The transformation $e^{B(\eta)}$ is the unitary operator
\begin{equation}\label{eq:T}
e^{B(\eta)} = \exp \Big( \frac{1}{2} \sum_{q \in P_H} \eta_q \left[ b_q^* b_{-q}^* - b_q b_{-q} \right] \Big)
\end{equation}
where $b_p = N^{-1}(N-\cN_+)^{1/2} \, a_p $ and $b^*_p =  N^{-1}a_p^* \, (N-\cN_+)^{1/2}$ are modified creation and annihilation operators which create or annihilate excitations, leaving the number of particles invariant. The coefficients $\eta_q$ are related to the solution of the scattering equation (\ref{zeroEnergySE}) and satisfy $|\eta_p| \leq C|p|^{-2}$, for $C>0$. The momenta in the sum in the exponent belong to the set $P_H= \{p\in \Lambda_+^*: |p|\geq \ell^{-\alpha}\}$, for parameters $\ell,\alpha>0$. Later we will fix a suitable $\alpha$ and choose $\ell$ small enough (but independent of $N$). We will exploit for example that  $ \sum_{q \in P_H} |\eta_q|^2\leq C \ell^{\a/2}$ is a small quantity.
Since $e^{B(\eta)}$ maps $\cF^{\leq N}_+$ back into itself, we can use it to define a new excitation Hamiltonian 
\begin{equation}\label{eq:cG} 
 \cG_{N,\ell} = e^{-B(\eta)} U_N H_N U_N^* e^{B(\eta)} : \cF_+^{\leq N} \to \cF_+^{\leq N}.
\end{equation}
The constant contributions in the new excitation Hamiltonian $\cG_{N,\ell}$ combine together to give $4\pi \frak{a}N$, i.e., the correlation structure introduced by $e^{B(\eta)}$ correctly resolves the leading order of the ground state energy.
Our goal is now to prove that for a suitable choice of $\a$ and $\ell$ small enough, there exist constants $C,c > 0$ such that 
\begin{equation}\label{eq:cGN-fin} 
\cG_{N,\ell} -4\pi \frak{a} N \geq c \cN_+ - C
\end{equation}
for all $N \in \bN$ sufficiently large.  The inequality in \eqref{eq:cGN-fin} clearly implies \eqref{eq:BEC} when evaluated on approximate ground states satisfying \eqref{eq:appGS}.

\emph{Step 3: Localization in Fock space.} To prove \eqref{eq:cGN-fin}, we divide the Fock space in two parts using the localization technique from \cite{LSo,LNSS}. We define $f,g: \bR \to [0;1]$ to be smooth functions with $f^2 (x) + g^2 (x)= 1$ for all $x \in \bR$. We assume that $f (x) = 0$ for $x > 1$ and $f (x) = 1$ for $x < 1/2$. We set $f_M = f (\cN_+ / M), g_M = g (\cN_+ / M)$. The function $f_M$ localizes therefore to Fock space sectors with $\cN_+\leq M$ and $g_M$ to sectors with $\cN_+\geq M/2$. We will chose at the end $M  = \ell^{3\alpha + \kappa} N$, for a suitable $\kappa > 0$. Using the properties of  $\cG_{N,\ell}$, we can prove the localization estimate
\begin{equation}\label{eq:cGN-loc} 
\cG_{N,\ell} =  f_M\, \cG_{N, \ell}\, f_M + g_M\, \cG_{N, \ell}\, g_M + \cE_{M}
\end{equation}
 for a $C > 0$ and a small error $\cE_{M}$ (in fact proportional to $M^{-2}$). We analyze now $f_M\, \cG_{N, \ell}\, f_M$ and $g_M\, \cG_{N, \ell}\, g_M$ separately. For the latter, we observe that there exists a constant $C > 0$ such that 
\begin{equation}\label{eq:gMGb}
g_M (\cG_{N,\ell} - 4\pi \frak{a} N ) g_M \geq C N g_M^2 \geq C \cN_+ g_M^2
\end{equation}
for all $N$ sufficiently large. The first inequality in \eqref{eq:gMGb} follows from a contradiction argument: if it was not true, we could find a sequence of states with high number of excitations and with the correct energy at leading order. Those would be approximate ground states as defined in \eqref{eq:appro-gs}, which therefore exhibit condensation in the zero momentum mode, contradicting the assumption $\cN_+\geq M/2= \ell^{3\alpha + \kappa} N/2$. The second inequality in \eqref{eq:gMGb} obviously follows from the first. This proves the first part of \eqref{eq:cGN-fin}. To conclude the proof of \eqref{eq:cGN-fin} it remains to prove for $f_M\, \cG_{N, \ell}\, f_M$ the bound analogous to \eqref{eq:gMGb}.

\emph{Step 4: Renormalizing cubic transformation}. In the last step we prove that for $\ell >0 $ small enough
\begin{equation}\label{eq:fMGN}
f_M (\cG_{N,\ell} - 4 \pi \frak{a} N) f_M \geq C \cN_+ f_M^2 - C \ell^{-3\alpha} f_M^2
\end{equation}
To achieve this, we define the operator $A : \cF_+^{\leq N} \to \cF_+^{\leq N}$ by 
\begin{equation}\label{eq:Aell1} A = \frac1{\sqrt N} \sum_{r\in P_{H}, v \in P_{L}} 
\eta_r \big[b^*_{r+v}a^*_{-r}a_v - \text{h.c.}\big],
\end{equation}
where we introduced the low--momentum set 
\[  P_{L} = \{p\in \Lambda_+^*: |p| \leq \ell^{-\beta}\} \]
for a parameter $0< \beta < \alpha$.
Then, for suitable restrictions on $\a$ and $\b$, there exists $\kappa > 0$ and a constant $C>0$ such that  
\begin{equation} \label{eq:propRnell}
\begin{split} 
\cR_{N,\ell} = e^{-A} \,\cG_{N,\ell}\,e^{A}\geq &\; 4\pi\mathfrak{a} N +  \big(1 - C\ell^{\kappa}\big) \cN_+   - C\ell^{-3\alpha}\cN_+^2/N   - C\ell^{-3\alpha}
\end{split}
\end{equation}	
for all $\ell$ small enough and $N$ large enough. The inequality \eqref{eq:fMGN} follows from \eqref{eq:propRnell} and the fact that conjugation with $e^A$ does not change significantly powers of the particle number operator, i.e.,
\begin{equation}\label{eq:gronA}
  e^{-A} (\cN_++1)^k e^{A} \leq C (\cN_+ +1)^k
\end{equation}
for all $\alpha > \b > 0$ and $N$ large enough. Equation \eqref{eq:gronA} can be proved using a Gronwall argument. Combining \eqref{eq:gMGb} and \eqref{eq:fMGN} with \eqref{eq:cGN-loc} we get to \eqref{eq:cGN-fin}.

\section{The excitation spectrum}

We discuss now the ideas in \cite{BBCS3} for proving Theorem \ref{thm:main}.

\emph{Step 1: Stronger bounds on $\cG_N$.} We define $\cG_N$ as in \eqref{eq:cG}, but in the definition of $B(\eta)$ (Eq. \eqref{eq:T}) we let the sum run over all momenta in $p\in\L^*_+$. We study now $\cG_N$ and determine it up to an error which vanishes for large $N$, getting
\begin{equation} \label{eq:deco2-GN}
\cG_N = C_{\cG_N} + \cQ_{\cG_N} + \cH_N + \cC_N  + \cE_{N},
\end{equation}
where $C_{\cG_N}$ is a constant contribution, $\cQ_{\cG_N}$ is quadratic, $\cH_N$ is the Hamiltonian \eqref{Hxspace} restricted to $\cF_+^{\leq N}$ and $\cC_N$ is the cubic operator
\begin{equation} \label{eq:cCN}
\cC_N = \frac{1}{\sqrt N} \sum_{\substack{p,q \in \L^*_+ \\ q \neq -p}} \widehat V(p/N)\,\Big[ b^*_{p+q} b^*_{-p} \big(  \text{cosh}(\eta_q)  b_q + \text{sinh}(\eta_q) b^*_{-q} \big) + \text{h.c.} \Big] \, . 
\end{equation}
The error $\cE_{N}$ satisfies
\begin{equation}\label{eq:error} 
\pm \cE_{N} \leq CN^{-1/4} \,  (\cH_N+\cN_+^2+1)(\cN_++1)\, .  
\end{equation}
The product on the right hand side is small on the states in which we are interested: the techniques we use to prove the bound for the number of excitations also allow us to prove bounds for the energy of excited particles $\cH_N$ and the products of $\cH_N$ with powers of $\cN_+$. This holds for states  $\xi_N = e^{-B(\eta)} U_N \psi_N\in \cF_+^{\leq N}$  with normalized $\psi_N \in L^2_s (\Lambda^N)$ belonging to the spectral subspace of $H_N$ with energies below $E_N + \zeta$, for some $\zeta > 0$. For such $\xi_N$, for any $k\in\mathbb N$ there exists a constant $C > 0$ such that 
\begin{equation}\label{eq:hpN} \langle \xi_N, (\cN_+ +1)^k (\cH_N+1) \xi_N \rangle \leq C (1 + \zeta^{k+1}) \, . \end{equation}

\emph{Step 2: Renormalize $\cQ_N$ through a cubic transformation.} We define a new cubic operator $\tilde A:\cF_+^{\leq N}\to\cF_+^{\leq N}$ by   
\begin{equation}\label{eq:defA}
        \begin{split}
     \tilde  A &= \frac1{\sqrt{N}} \sum_{\substack{ r\in \tilde P_H, v\in \tilde P_L }} \eta_r b^*_{r+v} b^*_{-r} \big[ \text{sinh}(\eta_v)b^*_{-v} +  \text{cosh}(\eta_v) b_{v}  \big] - \text{h.c.}\\
        \end{split}
\end{equation}
where $\tilde P_L = \{\, p \in \L_+^* : |p|\leq N^{1/2} \}$ corresponds to low momenta and $\tilde P_H =\L_+^* \setminus \tilde P_L$ to high momenta (by definition $r + v \not = 0$). The coefficients $\eta_p$ are the same as those used in the definition of the generalized Bogoliubov transformation $e^{B(\eta)}$ appearing in \eqref{eq:T}. We define the excitation Hamiltonian
\begin{equation}\label{eq:cJN-def}
\cJ_N := e^{-\tilde A} e^{-B(\eta)} U_N H_N U_N^* e^{B(\eta)} e^{\tilde A} = e^{-\tilde A} \mathcal{G}_N e^{\tilde A} : \cF_+^{\leq N} \to \cF_+^{\leq N}.
\end{equation}
We study now $\cJ_N $ and prove that
\begin{equation*}
\begin{split}
\cJ_N = C_{\cJ_N} + \cQ_{\cJ_N}  + \cV_N + \cE_{N},
\end{split}
\end{equation*}
where $\cE_{N}$ is an error term satisfying \eqref{eq:error} and $ \cV_N$ is the potential term of the Hamiltonian restricted to $\cF_+^{\leq N}$.
Conjugation with $e^{-\tilde A}$ cancels the cubic contribution \eqref{eq:cCN} and renormalizes the constant and quadratic contribution into new expressions $C_{\cJ_N}$ and $\cQ_{\cJ_N}$; there only the scattering length appears because all the instances of the interaction potential are corrected with the solution $f$ of the scattering equation \eqref{zeroEnergySE}.

\emph{Step 3: Diagonalization.} In the final step we act on $\cJ_N$ with a unitary transformation which approximately diagonalizes $\cQ_{\cJ_N}$. This is again a generalized Bogoliubov transformation $e^{B(\tau)}: \cF_+^{\leq N}\to \cF_+^{\leq N}$ with
        \begin{equation*}
        B(\tau):=\frac{1}{2}\sum_{p\in\Lambda^*_+}\tau_{p}\big(b^*_{-p}b^*_p-b_{-p}b_p\big) \, ,
 \end{equation*}
 where the coefficients $\tau_p \in \bR$ are suitably chosen so to approximately diagonalize $\cQ_{\cJ_N}$.
 The new excitation Hamiltonian
 \begin{equation*} 
\cM_N =  e^{-B(\tau)} \cJ_N e^{B(\tau)} = e^{-B(\tau)} e^{-\tilde A} e^{-B(\eta)}U_N H_N U_N^* e^{B(\eta)} e^{\tilde A} e^{B(\tau)}\end{equation*}
still leaves the space of excitations $\cF_+^{\leq N} $ invariant, and can be determined up to a small error $\cE_{N}$ to be given by
 \begin{equation}\label{eq:cor}
 \begin{split} \cM_{N} = &\;4\pi (N-1) \frak{a}+ e_\L \mathfrak{a}^2 + \frac{1}{2} \sum_{p \in \Lambda^*_+} \left[ - p^2  - 8\pi \frak{a}  + \sqrt{p^4 + 16 \pi \frak{a}  p^2} + \frac{(8\pi \frak{a} )^2}{2p^2} \right] \\ &+ \sum_{p \in \Lambda^*_+} \sqrt{p^4 + 16 \pi \frak{a}  p^2} \; a_p^* a_p + \cV_N + \cE_{N}. \end{split} \end{equation}
 The potential energy of excitations $ \cV_N$ is small on low-energy eigenspaces of the quadratic operator $\sum_{p \in \Lambda^*_+} \sqrt{p^4 + 16 \pi \frak{a} p^2} \; a_p^* a_p $. 
 Equations \eqref{1.groundstate} and \eqref{1.excitationSpectrum} follow now from \eqref{eq:cor} and the  min--max principle.

\section*{Acknowledgments}

C. Boccato acknowledges support from the European Research Council (ERC) under the programme Horizon 2020 (grant agreement 694227).

\end{document}